\documentclass[12pt,a4paper,titlepage]{article}
\usepackage{amsmath,amsthm,amsfonts}
\usepackage[active]{srcltx}
\usepackage{graphicx}
\begin{document}

\newcommand{\EF}{\mathcal{E}}
\newcommand{\abs}[1]{\left\lvert #1 \right\rvert}
\newcommand{\norm}[1]{\left\lVert #1 \right\rVert}
\newcommand{\ket}[1]{\lvert #1 \rangle}
\newcommand{\brkt}[2]{\langle #1 \vert #2 \rangle}
\newcommand{\braket}[3]{\langle #1\lvert #2\rvert #3\rangle}
\newcommand{\ketbra}[2]{\lvert #1 \rangle\langle #2 \rvert}
\newcommand{\schrod}{Schr\"odinger }
\newcommand{\bs}{\negthickspace}
\newtheorem{lemma}{Lemma}
\newtheorem*{thm}{Theorem}
\newtheorem{unbounded}{Theorem}
\parskip = 0.5\baselineskip

\begin{titlepage}
\begin{center}
{ \tt }\vspace{1cm}

\vspace{1cm}  {\huge \bf A Quantum Twin Paradox
\vspace{1cm}}

{ \tt }\vspace{1cm}

 {\large \bf }\vspace{1cm}

 { \tt by}\vspace{1cm}

 {\large \bf D. M. Ludwin}

 \vspace{0.1cm}
 {\small Department of Physics, Technion, Haifa 32000, Israel}

 \vspace{0.001cm}
 {\small \emph{ludwin@physics.technion.ac.il}}

\end{center}
\textbf{Abstract}\\
The Classical Twin Paradox is widely dealt in literature and neatly resolved. In addition, it is also well known that, when looking at two systems which are boosted relative to each other, the concept of the simultaneous effect of a quantum measurement in space-time causes some discrepancies in the cause-effect behavior. However, these discrepancies have been thought not to cause any apparent paradox except for violating the Free-Will postulate.
In this paper we suggest that using the local $t$ axis, all over space, as the axis in which the quantum measurement is thought to be simultaneous, we do reach a kind of true "Twin Paradox".
The resolution of this paradox requires the introduction of a global proper time into a covariant quantum theory.

\end{titlepage}
\pagenumbering{arabic}
\section{Introduction}

The Classical Twin Paradox is a thought experiment in special relativity, in which a twin makes a journey into space in a high-speed rocket and returns home to find that he has aged less than his identical twin who stayed on Earth. This result appears puzzling because each twin sees the other twin as traveling and so, according to a naive application of time dilation, each should paradoxically find the other to have aged more slowly. In fact, the result is not truly a paradox, since it can be resolved within the standard framework of special relativity by taking into account that the two twins don't really behave in a symmetrical manner - only one of them has undergone acceleration and deceleration. This effect has been verified experimentally using measurements of precise clocks flown in airplanes \cite{HK72} and satellites.

When we take the "Twin Paradox" into the world of the effect of a measurement in quantum mechanics, we get some discrepancies in the cause-effect behavior. Two inertial systems that move relative to each other with some constant velocity will disagree as to who first performed a measurement. In the case of two entangled particles, in which a measurement on one particle has a "simultaneous" effect on the other, this disagreement becomes a serious matter \cite{AA81}. Free-Will is enforced in nature using the fact that a single result of a measurement isn't totally bounded by physical laws, but rather has a statistical nature. However, if a measurement of an entangled state in a far location restricts the result of a measurement done on the other part of the entangled system, then only the first one to perform the measurement may be considered to have Free-Will.
Furthermore, if there is a disagreement concerning who truly measured first, it means that probably neither has true Free-Will, and that the entangled measurement both of them see might be determined by some kind of hidden variable arising from a statistical mechanism generated outside of these measurements and which is independent of the inertial system \cite{A95}.
This conflict between observers was dealt with by many authors (including Niels Bohr himself \cite{B35}) who have not been able to think of an apparatus for which there will be a difference in the physical results depending on who measured first \cite{A95}. They therefore concluded that there will be no true \emph{twin paradox} caused by this quantum nature of simultaneous measurement.

In this paper I shall show that in including a quantum eraser \emph{Delayed Choice} measurement, one can apparently achieve such a paradox.

Wheeler has presented the idea \cite{W78} that a quantum measurement always has a "delayed choice" nature, meaning that when a measurement is performed on a system it actually chooses a certain possible path in the history of the system, changing the entire state including the state of other subsystems which are entangled to it \cite{LB01}. In particular, a Which Way (WW) measurement will introduce orthogonality between the two main parts of the wavefunction, destroying the interference between these parts that otherwise might have been visualized. This WW measurement affects the whole historical path of the particle and enables or disables both parts of the wave function from interfering with each other in a "delayed choice" manner. Scully has shown in his famous "quantum eraser" experiment \cite{SZ97}, that even after the WW information has been marked in the system, it is still possible to erase this information and return to the original interference pattern.

In this paper, we shall present an apparatus for which we send a particle distribution through a Mach-Zehnder Interferometer and measure WW information on the path of the particles. It is clear that when measuring the interference pattern at the output of the interferometer, we shall have very different patterns depending on whether the WW information was erased before or after the interference pattern was measured.

A paradox arises if one can design the WW detector to be in a different inertial system with velocity $v$ relative to the Mach-Zehnder. In this case it can be arranged that, according to the "Instantaneous" idea of the observer in the WW inertial frame, the WW information is erased "before" the particles arrive at the detectors, while, according to his twin, in the Mach-Zehnder frame of reference, the erasing was done after the WW information effect on interference has already been observed in the lab. Both observers shall have a serious disagreement on the interference pattern that should actually appear.

The paradox that we get in this case, proves that the concept of spontaneous effect of quantum measurement can't be simplified to happen in the spatial axis 't', but requires a new definition of a \emph{global invariant time} to which different parts of the wave function are correlated \cite{HP73}.

\section{A quantum Eraser Experiment in a Mach-Zehnder Interferometer}

\begin{figure}
\begin{center}
  \includegraphics{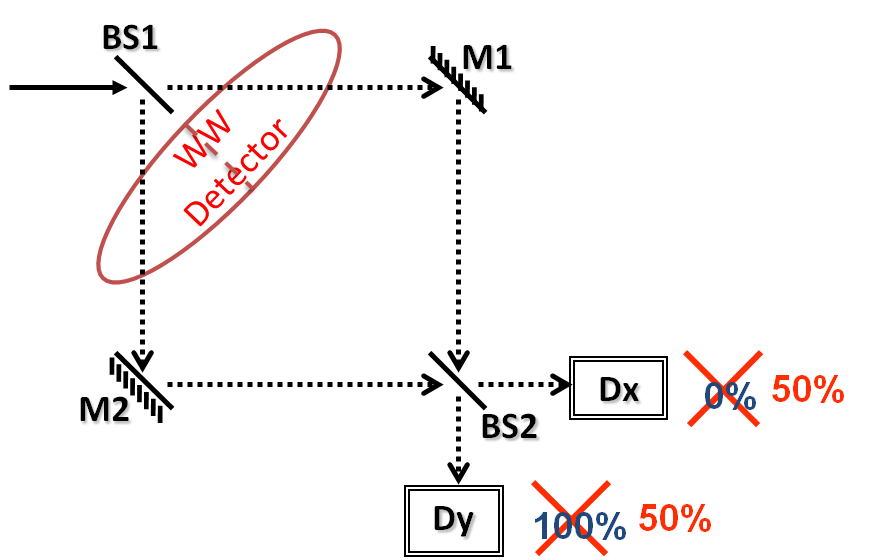}\\
  \end{center}
  \caption{Schematic Description of a Mach-Zehnder Setup with a WW detector. Without the WW information in the detector the output of the interferometer will be 0\% in Dx and 100\% in Dy. With WW information the distribution will be 50\%-50\%.}\label{WWMachZehnder}
\end{figure}

We shall look at an interference apparatus with the Mach-Zehnder Interferometer shown in figure
\ref{WWMachZehnder}. We can imagine a Mach-Zehnder interferometer for particles other than photons, in which the Beam-Splitters and mirrors (BS1, BS2, M1 and M2) are carried out using, for instance, Brag Scattering of the particles on a layer of atoms. We then take a bunch of particles and send them one by one through the interferometer. Each particle's wave function splits into two main parts: one, $|\Psi1\rangle$, is the part that passes through $BS1$ towards $M1$, and the other, $|\Psi2\rangle$, is the part that passes through $BS1$ towards $M2$.

We set the interferometer to have an interference pattern at the output such that all particles (100\%) shall be measured by a detector (Dy) placed in the $y$ direction while none (0\%) shall be measured by the detector (Dx) placed in the $x$ direction.

The interference between the two possible paths that the particle has taken is due to the fact that after interacting with $BS2$, the two paths which propagate in the same direction (for instance towards the Dx detector) have only a phase shift between them, caused by the fact that $\langle \Psi1|\Psi2\rangle$ is non vanishing and responsible for all the interference phenomena.

We now add a set of WW detectors to the interferometer, where each WW detector identifies which path a certain particle has taken. Since each WW detector is related to a specific particle, we can hereon treat each particle separately. Therefore, the WW detector has two states: one, $|WW1\rangle$, when the particle is identified as propagating towards $M1$ and therefore entangled to $|\Psi1\rangle$, and the other, $|WW2\rangle$, when the particle is identified as propagating towards $M2$ and therefore entangled to $|\Psi2\rangle$.

The total wavefunction of the particle entangled to the WW detector is therefore:
\begin{equation}\label{Eq. Total Wavefunction with WW}
    |\Psi\rangle=|\Psi1\rangle|WW1\rangle+|\Psi2\rangle|WW2\rangle
\end{equation}

An efficient WW detector is one that distinguishes correctly between the two WW states, and therefore can identify correctly the particle's path. In the WW Hilbert space, it means that the two WW states are orthogonal to each other such that: $\langle WW1|WW2\rangle=0$.

A WW detector will therefore cause the two parts of the particle's wavefunction to no longer interfere since including the WW states in the total Hilbert space will cause the interference elements, $\langle \Psi1|\Psi2\rangle\langle WW1|WW2\rangle$, to vanish. This will cause a 50\%-50\% distribution in both detectors at the output of the interferometer.

One can clearly see that if the WW information is deleted \textbf{before} the two parts of the wavefunction interfere, causing both states of the WW detector to overlap ($\langle WW1|WW2\rangle\approx1$). Then, the interference pattern caused by the element $\langle \Psi1|\Psi2\rangle\langle WW1|WW2\rangle$ and its C.C. will be recovered. If this is done for the whole set of WW detectors, all the particles will be measured in the Dy detector while none will be measured in the Dx detector. However, if the WW information is deleted only \textbf{after} the two parts of the wavefunction interfere, the particles have already been measured in the detectors with the WW information. In fact, some of the particles will have been definitely measured in the Dx detector, leaving the quantum erasing to be irrelevant.

This result of quantum erasing is of course by itself quite astonishing. It brings us to think of an apparatus where information concerning the act of erasing the WW information can be carried out faster than the speed of light between both parts of the interferometer. One can perform a statistical test at the Dx detector to figure out whether or not the erasure has been done in the WW detectors according to the percentage of particles found in the Dx detector, ~0\% or 50\% respectively.
In the following section, we shall show that a paradox can emerge since the interval between the WW information erasing act and the actual interference measurement could become space-like.

\section{A Quantum Twin Paradox in a Quantum Eraser apparatus.}

We shall now try to think of a WW detector which is in an inertial system with velocity $v$ relative to the Mach-Zehnder.
The WW information (for each particle separately) is collected when the WW detectors passed by the Mach-Zehnder interacting with a specific passing particle. Later, in the WW detector's frame of reference, the WW information is deleted from the detectors enabling the interference pattern to be recovered.

\begin{figure}
\begin{center}
  \includegraphics{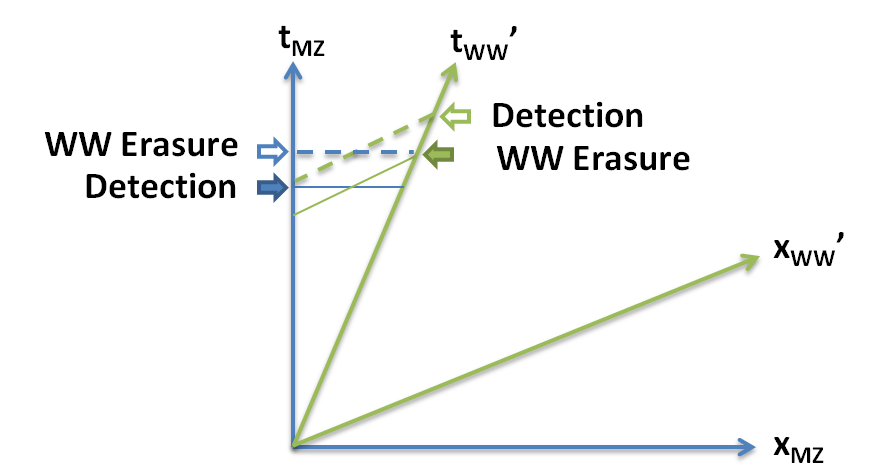}\\
  \end{center}
  \caption{A WW erasure has been done in system WW', which according to the instantaneous interpretation of the WW' system, which is parallel to the Xww' axis, it has been done before the detection in the MZ system. However, according to the MZ system, the WW erasure has been done after the detection.}\label{QTwinParadox}
\end{figure}

As is shown in figure \ref{QTwinParadox}, a paradox arises since according to the "Instantaneous" idea of the observer in the WW inertial frame, the WW information is erased "before" the particles arrive at the detectors, while, according to his "twin", in the Mach-Zehnder frame of reference, the erasing was done after the WW information effect on interference has already been observed in the lab.
Both twins will seriously disagree on whether there are any particles measured in the Dx detector or not.

\section{Discussion}
The paradox described above proves that the concept of the spontaneous effect of a quantum measurement can't be simplified to happening in the spatial axis 't', but rather needs a new definition of a \emph{global invariant time} which different parts of the wave function are correlated with each other according to.

The straightforward candidate for a \emph{global time} will be Einstein's \emph{proper time}.
We can think of the effect of time dilation in special relativity quite differently. Suppose that the true correlation between two parts of an entangled system is \emph{proper time}. This means that \emph{proper time} is the actual beating clock that gives the system its evolution.
It also means that in the case of true entangled twin particles that went on some journey, we can't actually have them interfere quantum mechanically, unless they meet at the same \emph{proper time}, with a small enough interval in space-time which can be allowed for by the uncertainty in the 4D location of the particles. Understanding this, the effect of simultaneous measurement in quantum mechanics will be according to \emph{proper time} and all the paradoxes vanish.

In a later paper, we shall show that \emph{proper time} isn't exactly the correct candidate to correlate the wave function, but it is quite close.


\newpage
\begin{thebibliography}{99}

\bibitem{HK72}
J. Hafele, R. Keating, "Around the world atomic clocks:predicted relativistic time gains", Science \textbf{177}, 166 (1972).
\bibitem{AA81}
Y. Aharonov and D. Z. Albert, Phys. Rev. D \textbf{24}, 359 (1981)
\bibitem{A95}
A. Peres, "Quantum Theory: Concepts and Methods", Kluwner Academic Publishers: \emph{Fundamental Theories of Physics} Volume \textbf{72} (1995), Ch.6 "Nonlocality Vs. free will"
\bibitem{B35}
N. Bohr, Can Quantum-Mechanical Description of Physical Reality be Considered Complete?", Physical Review, \textbf{48}, 696 (1935)
\bibitem{W78}
J. A. Wheeler, "The 'past' and the 'delayed choice' double-slit experiment", in A.R. Marlow, ed., \emph{Mathematical Foundations of Quantum Theory} (Academic Press, New York 1978), 9-48
\bibitem{LB01}
The Role of Momentum Transfer in Welcher-Weg Experiments, D. M. Ludwin and Y. Ben-Aryeh, Foundations of Physics Letters \textbf{14(6)}, 519-528 (2001).
\bibitem{SZ97}
M.O. Scully and M. S. Zubairy, "Quantum Optics", (Cambridge University Press 1997), 568
\bibitem{HP73}
Relativistic dynamics, L.P. Horwitz and C. Piron, Helv. Phys. Acta \textbf{46}, 316 (1973).


\end{thebibliography}
\end{document}